\documentclass[12pt,a4paper,final]{iopart}

\usepackage{iopams}  
\usepackage{graphicx}
\usepackage{verbatim}
\usepackage{textcomp,bbm}
\usepackage[breaklinks=true,colorlinks=true,linkcolor=blue,urlcolor=blue,citecolor=blue]{hyperref}

\newcommand{\F}{\mathcal{F}}

\begin{document}

\title[Transverse-mode coupling and diffraction loss in microcavities]{Transverse-mode coupling and diffraction loss in tunable Fabry-P\'erot microcavities}

\author{Julia Benedikter$^{1,2}$, Thomas H{\"u}mmer$^{1,2}$, Matthias Mader$^{1,2}$, Benedikt Schlederer$^{1,2}$, Jakob Reichel$^{3}$, Theodor W. H\"ansch$^{1,2}$, David Hunger$^{1,2}$}
\address{$^1$Fakult{\"a}t f{\"u}r Physik, Ludwig-Maximilians-Universit{\"a}t, Schellingstra{\ss}e~4, 80799~M{\"u}nchen, Germany}
\address{$^2$Max-Planck-Institut f{\"u}r Quantenoptik,  Hans-Kopfermann-Str.~1, 85748~Garching, Germany}
\address{$^3$Laboratoire Kastler Brossel, ENS/UPMC-Paris 6/CNRS, 24 rue Lhomond, 75005~Paris, France}
\ead{david.hunger@physik.lmu.de}

\begin{abstract}
We report on measurements and modeling of the mode structure of tunable Fabry-P\'erot optical microcavities with imperfect mirrors. We find that non-spherical mirror shape and finite mirror size leave the fundamental mode mostly unaffected, but lead to loss, mode deformation, and shifted resonance frequencies at particular mirror separations. For small mirror diameters, the useful cavity length is limited to values significantly below the expected stability range. We explain the observations by resonant coupling between different transverse modes of the cavity and mode-dependent diffraction loss. A model based on resonant state expansion that takes into account the measured mirror profile can reproduce the measurements and identify the parameter regime where detrimental effects of mode mixing are avoided.

\end{abstract}

\vspace{2pc}
\maketitle

\section{Introduction}

Fabry-P\'erot optical microcavities built from micro-machined concave mirrors \cite{Hunger10b,Hunger12,Trupke05,Muller10,Dolan10,Greuter14,Takahashi14} offer a powerful combination of small mode cross section, high finesse, and open access. This has proven to be beneficial for experiments covering a broad range of topics, including cavity quantum electrodynamics with cold atoms \cite{Colombe07,Trupke07}, ions \cite{Steiner13,Brandstaetter13}, and solid-state-based emitters \cite{Toninelli10,Muller09,Barbour11,Di12,Albrecht13,Kaupp13,MiguelSanchez13}, as well as cavity optomechanics \cite{Favero09,Stapfner13,FlowersJacobs12} and scanning cavity microscopy \cite{Mader14}.
Various techniques have been developed to produce concave, near-spherical profiles as mirror substrates, including CO$_2$ laser machining \cite{Hunger12,Muller09,Greuter14,Uphoff15,Takahashi14}, chemical etching \cite{Trupke05,Biedermann10}, focused ion beam milling \cite{Dolan10,Albrecht14}, and thermal reflow \cite{Cui06,Roy11}.
A small cavity mode cross section is achieved by realizing microscopic surface profiles with radii of curvature $r_c\sim 5 - 500\,\mu$m and profile diameters typically a factor $2 - 10$ smaller. In this regime, the extent of the cavity mode can be comparable to the effective mirror diameter, and the finite mirror size becomes relevant. Furthermore, the different fabrication processes typically yield profiles that deviate from a spherical shape, and excessive surface roughness may be present. In addition, coating defects and particles on the mirror surface can disturb cavity performance under real conditions. Overall, the mode structure of open-access microcavities will be affected by the details of the mirrors, and an in-depth understanding of the relation between mirror imperfections and cavity performance is required for the successful application and the potential improvement of such resonators.

In this work we study the consequences of finite mirror size and non-ideal shape on the performance of laser-machined, fiber-based Fabry-P\'erot microcavities \cite{Hunger10b}. Their mirrors are characterized by surface profiles with low microroughness in the range of $1 - 2\,$\AA, a near-spherical central part, and an overall shape that is well approximated by a Gaussian.

We perform measurements of the cavity transmission and finesse across the entire stability range for several cavities.
We find that for short mirror separation, the cavities are mostly immune to mirror imperfections, and the fundamental cavity mode closely resembles a Gaussian mode. However, at particular mirror separations, the cavity shows a significantly reduced finesse, and the performance depends for example on the precise laser wavelength. Furthermore, for small mirror size, we observe that the distance range where the finesse remains high is significantly smaller than the stability range expected from the mirror radius of curvature. 

We accurately reproduce these observations with a model based on resonant state expansion \cite{Klaassen05,Kleckner10, Muljarov10}, where we take into account the measured mirror profile.
The model shows that the observed behavior can be consistently explained by (near-) resonant coupling between different transverse modes of the cavity, caused by the non-ideal shape and finite size of the mirrors. The admixture of higher order modes, which suffer from diffraction loss due to their larger size, introduces loss to the fundamental mode. Based on these results, we identify the parameter regime for Gaussian-shaped mirrors where detrimental effects of mode mixing remain negligible.

\begin{figure*}
	\centering
	\includegraphics[width=\textwidth]{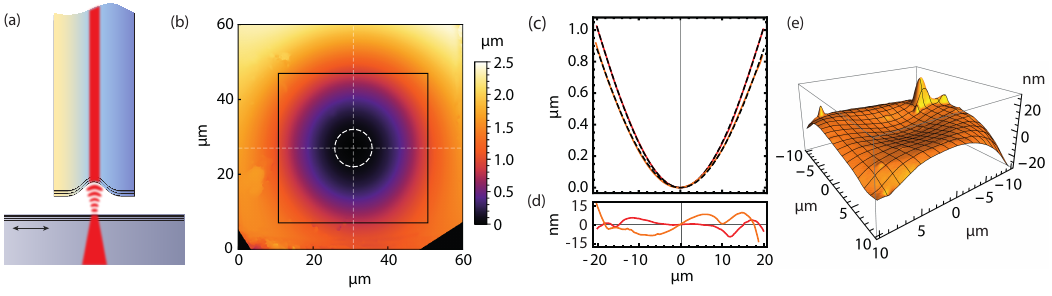}
	\caption{\label{fig:Mirror} (a) Schematic setup showing the machined fiber, the plane mirror, and the optical mode. (b) Mirror profile as measured with a white-light interferometer. Dashed circle: typical mode size ($ 1/e^2 $-diameter) on the mirror. Black square: Simulation area. (c) Profile sections (as indicated with dashed gray lines in (b)). Red (Orange): cut along x (y) -direction. Dashed lines: Gaussian fit. (d) Residuals of the fits in (c). (e) Residual of a 2D parabolic fit to the region of the profile typically covered by the cavity mode.
	}
\end{figure*}

\section{Fiber-based microcavity}

The cavity design is depicted schematically in Fig.\ \ref{fig:Mirror} (a): The resonator consists of a curved micromirror machined on the end-facet of a single mode optical fiber and a macroscopic plane mirror. Both the commercial plane mirror substrate and the fiber surface are coated with a highly reflective dielectric coating for a center wavelength of 780\,nm, where a finesse of $ \F\approx 60 000 $ can be reached. In this configuration, the planar mirror serves as a near-ideal reference mirror, which permits us to study the effects of the micromirror alone. The light of a grating-stabilized diode laser is coupled into the cavity through the fiber, and light transmitted through the plane mirror is collimated and detected with an avalanche photo diode. The whole stability range of the cavity can be covered with sub-nanometer resolution with a piezo step drive linear positioning stage (PI LPS-24), onto which the fiber is mounted. In addition, a shear piezo crystal is used for scanning the cavity length over the resonance. The plane mirror can be laterally scanned with an XY piezo table (PI P-541.2SL) over one hundred micrometers. A mirror mount allows for angular alignment of the cavity.

A white-light interferometric image of the laser-machined depression on the fiber surface is shown in Fig.\ \ref{fig:Mirror} (b). The image is taken with a home-built instrument with a lateral resolution of 560\,nm and a vertical resolution of 0.1\,nm (rms). The dashed white circle illustrates the $ 1/e^2 $-diameter of the fundamental mode for a mirror separation $ d \approx r_c/4$, the black square shows the area used for the simulation.
A Gaussian fit to the surface data (lines in Fig.\ \ref{fig:Mirror} (c)) shows good agreement, and only nanometer-scale deviations can be seen from the residuals (Fig.\ \ref{fig:Mirror} (d)). The central part of the profile can be well approximated by a parabola. Figure \ref{fig:Mirror} (e) shows the residual of a two-dimensional parabolic fit to the data. Certain localized imperfections are present on the fiber surface, as well as an overall shape deviation.

Figure \ref{fig:Mirror} (c) also illustrates that the profile is not rotationally symmetric but rather has elliptical contour lines. This leads to Hermite-Gauss modes to closely resemble the eigenmodes of the cavity, and to a splitting of higher-order transverse modes of equal order. Additionally, the ellipticity splits each cavity resonance into a linear polarization doublet \cite{Uphoff15}. We use polarization optics before the fiber to select one of the modes for evaluation. The surface shown in Fig.\ \ref{fig:Mirror} (b) is rotated such that the principal axes of the profile coincide with the coordinate axes. 
The minimal radius of curvature in the center is found to be $ r_c^{(x)} = 161\,\mu$m in $x-$ and $ r_c^{(y)} = 201\,\mu$m in $y-$direction. This is in the range of typical values for laser machined mirror profiles.

\section{Experimental Results}

We study the cavity performance by measuring the finesse for each accessible axial mode order. To ensure that local variations of the mirror coating of the planar mirror do not influence the result, we determine the finesse at 25 different positions within an area of $30\times 30\,\mu$m on the planar mirror and evaluate the most prevalent value. Diffraction loss, which arises as result of mode mixing, leads to a decreased finesse according to
\begin{equation}
\F = \frac{2\pi}{T+A+D},\label{finesse}
\end{equation}
where $ T $ denotes the total transmission of both mirrors, $ A $ the total absorption loss, and $D$ the diffraction loss due to the micromirror. In our experiment $ T+A \approx 100 $~ppm.

\begin{figure}
	\centering
	\includegraphics{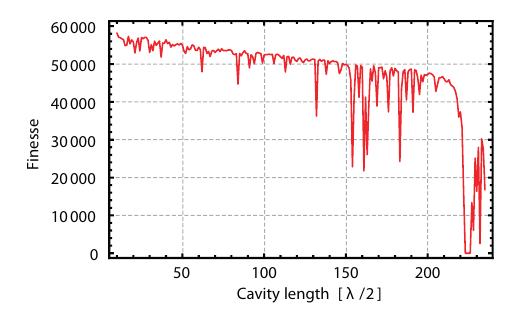}
	\caption{\label{fig:Finesse} Cavity finesse as a function of axial mode order determined from measurements at 25 different positions within an area of $30\times 30\,\mu$m on a large planar mirror and taking the most prevalent values. Wavelength: 780~nm.
	}
\end{figure}

A typical measurement of the finesse of the fundamental mode as a function of mirror separation is shown in Fig.\ \ref{fig:Finesse}. To obtain the quality factor $ Q $ of the cavity, we imprint sidebands as frequency markers using an EOM. The exact cavity length needed for determining the finesse from $ Q $ is inferred from the transmission spectra of two lasers of known wavelength. In such measurements, we typically observe three different regimes:
For small mirror separation $ d\lesssim r_c^{(x)}/4 $, (axial mode number $ q<130 $ for the measurement shown), the finesse stays approximately constant with only a slight overall decline. %In this regime, the transverse modes are closely spaced, and only very 
For intermediate mirror separations, individual axial mode numbers $ q $ show large additional loss.
For mirror separations $d\gtrsim r_c^{(x)}/2$, an abrupt drop of the cavity transmission and finesse is observed, with few moderately-working mode orders appearing for larger $d$.

We observe that increased loss appears mainly when higher-order transverse modes become degenerate with the fundamental mode. Therefore it is instructive to study cavity transmission spectra for different mirror separations. A typical spectrum of a cavity at high finesse is shown in Fig. 3 (a), where we probe the cavity with a narrow-band laser and tune the mirror separation across one free spectral range. To map out the mode profile, we raster-scan a nanoparticle placed on the large mirror using the cavity mode and evaluate the introduced loss from the cavity transmission of each of the modes \cite{Mader14}. The mode shapes clearly resemble those of Hermite-Gaussian modes orientated along the principal axes of the mirror profile. For a given mode with transverse mode order $ (m,n) $ and axial mode order $ q $, the cavity resonance frequencies for a spherical mirror cavity are given by

\begin{equation}
\nu_{qmn} = \frac{c}{2d} \left( \,q +\frac{1/2+m}{\pi}\xi^x + \frac{1/2+n}{\pi}\xi^y \right),
\end{equation}
where $\xi^{(x,y)}=\arccos\sqrt{1-d/r_c^{(x,y)}}$ is the Gouy phase.
Note that the degeneracy of modes with the same transverse mode order $ m+n $ is lifted by the ellipticity of the profile, leading to families of modes with $ m+n+1 $ members. In this regard, the mirror ellipticity is useful since it allows to study the impact of the modes separately.
Figure 3 (b) shows spectra like the one in (a) as a function of cavity length for 856\,nm probe light, where $ \F \approx 1200$ to improve the visibility of the resonances. 
The logarithmic color scale is set for each cavity length to optimize the signal-to-background ratio. 
With increasing mirror separation, the spacings between the transverse modes increase and eventually, higher-order modes can become resonant with the next fundamental mode. This is given when the differential Gouy phase fulfills
\begin{equation}
m\xi^x + n\xi^y = j\pi
\end{equation}
where $j$ is an integer.

\begin{figure}
	\centering
	\includegraphics{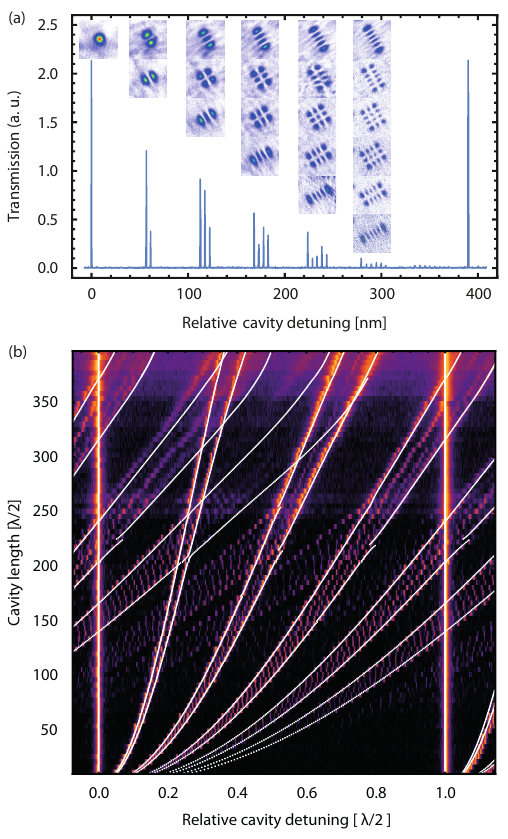}
	\caption{\label{fig:Spectra} (a) Cavity transmission spectrum as a function of relative cavity detuning covering one free spectral range, where $ \F \approx 60.000 $. The insets show the mode functions as measured by scanning cavity microscopy. Wavelength: 780~nm. (b) Cavity transmission spectra as a function of axial mode order $q$ for $ \F \approx 1200 $. Logarithmic color scale. Wavelength: 856~nm. White: model for modes $ (0,m+n) $ and $ (m+n,0) $.
	}
\end{figure}

When evaluating the measured transverse modes, we find that their frequencies deviate from the spectrum given by Eq.\ 2, and Eq.\ 3 fails to predict the positions of the observed resonances. An accurate description is possible by the model discussed in the next section.
The white lines in Fig.\ \ref{fig:Spectra} (b) show the predicted resonances of the modes $ (0,m+n) $ and $ (m+n,0) $ for the lowest few mode orders $ m+n $.

\begin{figure}
	\centering
	\includegraphics{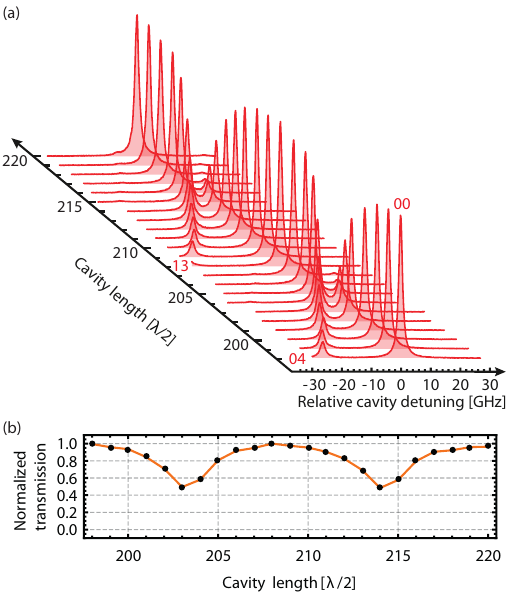}
	\caption{\label{fig:Transmission} (a) Measured cavity transmission spectra as a function of relative cavity detuning. Wavelength: 856~nm. Avoided crossings between the fundamental mode and the fourth order modes 04 and 13 are visible. (b) Maximal transmission of each of the spectra shown in (a), normalized to the largest value.
	}
\end{figure}

A closer look at the crossover of modes 04 and 13 with the fundamental mode is taken as an example for the typical avoided crossing behavior found when mode coupling occurs. Fig.\ 4 (a) shows spectra covering the region around the fundamental mode for every axial mode order from $q = 198$ to $q = 220$ for $ \mathcal{F}\approx 1200$.
Close to resonance, the transmission of the fundamental mode decreases (see Fig.\ 4 (b)), while the transmission of the higher order mode increases until both have approximately equal height and minimal separation at the point of resonant coupling. The coupling is accompanied with an increased linewidth and thus a reduced finesse. At the anticrossing between the 00 and 04 mode, we observe a mode splitting of 8.6 GHz. We have modeled the coupling of the particular mode pair (see below) and find a value of 8.6 GHz, matching the measurement within errors.

\section{Modeling}

For cavity mirrors where the surface profile can be treated as a perturbation of a spherical shape, an effective approach is to describe the real eigenmodes $ \Psi_i $ as a series expansion of Hermite-Gauss modes $ \Phi_k $,
\begin{equation}
\Psi_i = \sum_k c_{ik} \Phi_k, \ k = (m,n). \label{eq:expansion}
\end{equation}
% \ c_k \in \mathbb{C},
Following the approach of Kleckner et al.\ \cite{Kleckner10}, we determine the new eigenmodes and the corresponding resonance frequency and loss. 
Introducing the mode-mixing matrix $M$, which accounts for the change a mode undergoes during one round trip through the cavity, the task reduces to an eigenvalue problem
\begin{equation}
\gamma_i \Psi_i = M \Psi_i.
\end{equation}
The mixing matrix $ M_{k,k'} = \mathrm{exp}(-4i\pi d/\lambda) \mathbbm{1}_{k,k'} B_{k,k'}$ has elements given by mode overlap integrals taken over the finite extent of the micromirror,
\begin{equation}
B_{k,k'}=\left.\int_{-x_0}^{x_0}\int_{-y_0}^{y_0}\Phi^-_k\Phi^{+*}_{k'}e^{-4i\pi\Delta(x,y)/\lambda} dx dy \right|_{z=z_m}.
\end{equation}
Here, $(x_0,y_0)$ denote the extent of the mirror, $\pm$ indicates the sign of the phase factor of $\Phi_k$, $\Delta(x,y)$ is the deviation of the mirror profile from a planar surface, and $z_m$ is the location of the micromirror on the optical axis. We assume that the respective expression for the planar mirror is an identity matrix.

Using the Hermite-Gauss modes for the expansion implies the paraxial approximation, where the isophase surface is parabolic (with some deviation due to the Gouy phase) rather than spherical. However, the paraxial approximation does not hold at large separation from the optical axis where the two shapes differ. In fact, when including non-paraxial terms, one finds that a spherical geometry is indeed the most desirable \cite{Laabs99}. For a spherical mirror which covers an entire half-space, $ B_{k,k'} = \mathbbm{1}_{k,k'}$, and the eigenvalues $ \gamma_i $ corresponding to the eigenmodes $ \Psi_i $ are unity. As soon as $ \Delta(x,y) $ deviates from spherical or $x_0,y_0$ is finite, $M$ has off-diagonal elements and transverse-mode mixing occurs.

\begin{figure*}
	\centering
	\includegraphics[width=\textwidth]{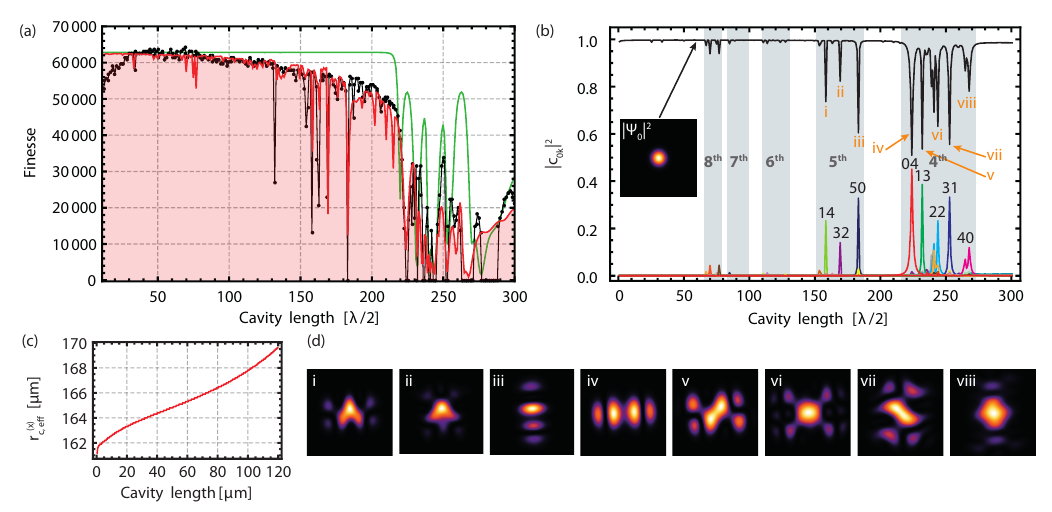}
	\caption{\label{fig:Simulation} (a) Data points: Cavity finesse extracted from 22500 measurements for each datapoint. Wavelength: 780~nm. Red: Simulation of the cavity finesse as a function of the mirror separation for a measured surface profile assuming mirror transmission and absorption to be 100\,ppm. Green: Simulation for a Gaussian profile fitted to the measured surface. (b) Composition of the ground mode $ \Psi_0 $ from Hermite-Gaussian modes $ \Phi_k $; contributions $ \left |c_{0k}\right|^2 $ are shown. Black: $ \left |c_{00}\right|^2 $. Black numbers denote the transverse mode order of other important contributions. Shaded areas show impact zones of specified mode orders. Inset:  $ \left |\Psi_0\right|^2 $ for a region of low coupling ($q=60$). (c) Effective radius of curvature $r_{c,\mathrm{eff}}^{(x)}$ sensed by the fundamental cavity mode as a function of cavity length. (d) Examples of mode shapes of the fundamental mode $ \left |\Psi_0\right|^2 $ for selected mirror separations indicated by Roman numbers in (b) exhibiting large coupling. The edge length is $40\,\mu$m.
	}
\end{figure*}

For an accurate treatment of our experiment, we use the measured surface profile (Fig. 1(b)) for $\Delta(x,y)$.
To find a suitable basis set $\Phi_k$ for each mirror separation, we numerically maximize $\left |M_{0,0}\right|$ by varying the mode waist $w_0$ of $\Phi_{00}$. For a given mirror separation, the obtained optimal $w_0$ corresponds to an effective radius of curvature $r_{c,\mathrm{eff}}$ of a spherical mirror. The result for the profile investigated here is displayed in Fig.\ \ref{fig:Simulation} (c), showing that $r_{c,\mathrm{eff}}^{(x)}$ is larger than $ r_c^{(x)} = 161\,\mu$m and increases with $d$. 
Consequently, the stability range is expected to extend beyond the limit of $ d=r_c $. 

The diffraction loss $D_i $ of a mode $ \Psi_i $ can be directly obtained from the corresponding eigenvalue $ \gamma_i $,
\begin{equation}
D_i = 1-|\gamma_i|^2.
\end{equation}
Inserting this into Eq.\ \ref{finesse} gives the finesse of this mode.

The obtained finesse of the fundamental mode using the profile shown in Fig.\ 1 (b) is plotted in red in Fig.\ 5 (a), where we use the measured mirror transmission and loss. For direct comparison, we show a measurement of the cavity finesse as obtained from the resonant cavity transmission ($ T_c \propto \F^2 $) at a wavelength of 780\,nm. We measure the transmission rather than the finesse here because we cannot determine the finesse reliably under mode mixing conditions as well as for low transmission. The rise of the finesse for short cavities can be attributed to a systematic error of the measurement: We use an iris aperture to suppress the transmission of higher order modes, which leads to clipping loss for short cavity lengths where the cavity mode radius is smaller and its divergence larger. The overall shape, the position of localized finesse dips, and the decrease around $q=220$ can be reproduced by the simulation with a high level of detail. However, to match the data, the lateral size of the mirror profile had to be rescaled by about 2.5\% for the simulation. The same correction has been made for the simulation of the spectrum shown in Fig.\ 3, where the normalized phase of the eigenvalues $ \gamma_i $ is plotted. The mismatch might result from a calibration uncertainty of the interferometric surface reconstruction.
The localized finesse dips correspond to narrow mode resonances involving high mode orders (see below), which are not resolved by the $\lambda/2$-discrete sampling. The resonance condition furthermore depends on the exact probing wavelength and on the dispersive mirror properties which vary spatially. The finesse values at the dips are thus somewhat arbitrary, and both measurement and simulation may miss particular resonances. To capture the typical behavior in the measurement, we have therefore measured at 22500 positions on a $ 30 \times 30\,\mu$m area of the plane mirror and take the most prevalent value for each data point shown.
%We note that in our model we do not take the vectorial character of the light field into account \cite{Uphoff15}, which together with additional non-paraxial corrections of the mode frequencies \cite{Yu84} is estimated to lead to relative frequency changes $\lesssim 10^{-3}$ for our cavity geometry. The associated polarization mode splitting furthermore leads to a polarization dependence of the mode mixing behavior.
%The exact position and depth of the simulated finesse dips highly depend on the wavelength and the chosen sampling lengths as the mode coupling impact zone for such high finesse is smaller than the distance between subsequent sampling points ($ \lambda /2 $). A quantitative prediction of the dip depth and the occurrence of dips at small mirror separation is therefore not possible.

The computed eigenvectors contain information about the composition of the system's eigenmodes from Hermite-Gaussian modes according to Eq.\ \ref{eq:expansion}. In Fig.\ 5 (b), the coefficients $ |c_{0k}|^2 $ giving the contributions to the fundamental mode are plotted as a function of mirror separation. The Gaussian mode $ \Phi_{00} $ is clearly the dominant one, and for most cavity lengths, the ground mode shows negligible deviation from it (see inset in Fig.\ \ref{fig:Simulation} (b)). However, for certain distinct mirror separations where resonant mode mixing occurs, higher order modes can have significant contributions and lead to a severe distortion of the fundamental mode (Fig.\ 5(d)). The larger spatial extent of higher-order modes with $w_k\approx w_0\sqrt{k+1}$ causes larger diffraction loss, from which also the fundamental mode suffers under coupling conditions. Notably, the locations of high loss and strong mode mixing do mostly but not necessarily coincide (see below and \cite{Kleckner10}).

Regions of impact of certain mode families (shaded areas) cover a significant fraction of the stability range. Still, for applications where the exact mirror separation is not essential, extended regions of negligible mode mixing remain. The different influence of e.g.\ mode orders 4 and 5 can be attributed to the larger values of $B_{0,k}$ for even modes due to the symmetric mirror profile and modes with smaller mode index. Also, for larger mode index differences, the differential Gouy phase evolves faster and the resonance condition is sharper. The coupling strength can be directly inferred from the mode splitting at an avoided crossing.

It is instructive to compare the results with a calculation for a profile obtained from a Gaussian fit to the measured fiber surface (green solid line in Fig.\ref{fig:Simulation} (a)). The smooth surface does not lead to the overall weak decline for increasing mirror separation, and the sharp features at intermediate $d$ are missing. Yet, the finesse decrease around $d = r_c/2$, which effectively limits the stability range, is reproduced. The difference can be explained by the presence of additional (and in particular asymmetric) surface deviations with mostly larger spatial frequencies and particle-like features in the measured profile. High spatial frequencies couple the fundamental mode to many transverse modes with large mode index, causing a smooth finesse decrease and significant resonant mixing for particular modes. 

We note that in our model we do not take the vectorial character of the light field into account \cite{Uphoff15}, which together with additional non-paraxial corrections of the mode frequencies \cite{Yu84} is estimated to lead to relative frequency changes $\lesssim 10^{-3}$ for our cavity geometry. The associated polarization mode splitting furthermore leads to a polarization dependence of the mode mixing behavior, which we observe in our experiments.

\begin{figure*}
	\centering
	\includegraphics[width=\textwidth]{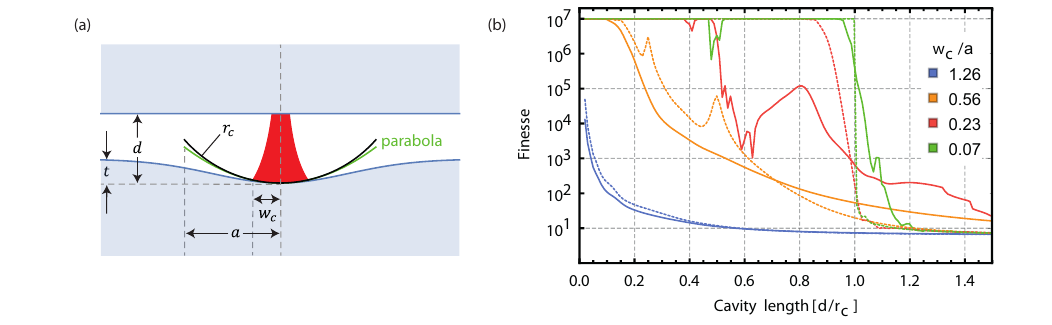}
	\caption{\label{fig:6} 
		(a) Schematic of the cavity showing the relevant parameters. (b) Simulated finesse for different mirror profile size at fixed radius of curvature $ r_c $ using various ratios of $ w_c/a $ defined at $d=r_c/2$. Solid lines: Gaussian profile. Dashed lines: parabolic profile. We assume additional loss of $ \pi\times10^{-7} $ for each mirror.
	}
\end{figure*}

The observed behavior is not limited to the particular parameters used in our experiment, but is a general property related to the profile shape and size. Considering a Gaussian profile with $1/e$ radius $a$, depth $t$, and $r_c = a^2/(2t)$, and assuming a cavity with $d=r_c/2$ where the mode radius on the curved mirror is $w_c = \sqrt{\lambda r_c /\pi}$, one finds that the relative mode size depends only on the profile depth for a given wavelength, $w_c/a = \sqrt{\lambda/(2\pi t)}$. The relevant quantities are visualized in Fig.\ \ref{fig:6} (a). We perform simulations for a profile with fixed $r_c$ and ellipticity $\epsilon = \sqrt{1-r_c^{(x)}/r_c^{(y)}} = 0.26$ \cite{Uphoff15} and vary $w_c/a$. The resulting finesse is shown in Fig.\ \ref{fig:6} (b). While profiles as small as $a = 2\, w_c$ already achieve performance not limited by diffraction for small mirror separation, it requires a profile radius $a > 4\, w_c$ to extend this range to $d = r_c/2$ and $a > 10\, w_c$ to avoid mode mixing over the entire stability range. For comparison, we also perform simulations for a rotationally symmetric parabolic profile with $ r_c=r_c^{(x)} $ and an edge length of $ 2a $. While the overall behavior is similar, the calculation for $w_c = 0.56\, a$ shows that resonant transverse-mode mixing can also lead to a reduction of diffraction loss \cite{Kleckner10}. This can be understood by the destructive interference between the fundamental and the higher-order mode at the outer part of the mode, reducing the effective mode size \cite{Mader14}.

For obtaining the presented data, modes up to order $ m+n=20 $ are included in the calculation. Using more modes does not significantly alter the results and strongly increases the computation time, which grows approximately as $ (m+n)^4 $. The required minimal size of the area used for the simulation depends on the profile details. For the example shown above, we have tested different sizes to confirm that the area chosen provides sufficient accuracy of the simulation results and that no significant dependence on the size is present at this scale. The pixel size is chosen such that the features of the highest mode order are still well resolved. For a fiber profile of $ 400\times 400 $ pixels covering an area of $40\times 40\,\mu$m$^2$, the simulation of 300 mirror separations could be conducted with a personal computer within a few hours \footnote{We are happy to provide the code.}.

\section{Conclusion}

Our results comprise an extensive analysis of the performance of optical cavities with non-ideal mirrors. We have shown that localized reduction in cavity finesse, frequency shifts, and mode shape distortions are the consequences of non-ideal mirror shape and finite size. The behavior can be accurately modeled with a mode expansion method when using the measured mirror profile as an input.
The demonstrated approach provides a powerful tool for analyzing a given cavity geometry and for predicting cavity performance. This is particularly helpful for experiments where minimal mode volume and ultimate small mirror profiles are desired, as well as for cold atom, ion trap, and cavity optomechanics experiments, where larger mirror separations in combination with small radii of curvature are beneficial. The calculations provide improved accuracy for the determination of emitter-cavity coupling strength as well as detailed information about possible sample-induced scattering and loss. Finally, the approach offers an efficient route for the design of novel cavity geometries with non-trivial properties, such as single-transverse-mode operation \cite{Kuznetsov04,Ferdous14}, where higher order modes can be suppressed without significantly affecting the fundamental mode e.g.\ to improve spectral filtering, or mode imaging \cite{Weitenberg11}, where a cavity mode is designed to avoid a scatterer to reduce loss.
This opens the potential for light modes to be individually tailored for specific applications.

\section*{Acknowledgments}

We thank Hanno Kaupp, Christian Deutsch, and Raphael Franz for support, and Jason Smith and Aurelien Trichet for helpful discussions.
This work has been partially funded by the Excellence cluster Nano Systems Initiative Munich (NIM) and by the European Union's Seventh Framework Programme (FP7) under grant agreement No.\ 618078 (WASPS).

\end{document}